\documentclass[sigplan]{acmart}
\renewcommand\footnotetextcopyrightpermission[1]{} 
\usepackage{booktabs} 

\usepackage{tikz}
\usetikzlibrary{shadows}
\usepackage{pgfplots}
\pgfplotsset{compat=1.10}

\usepackage{colortbl}
\definecolor{Gray}{gray}{0.90}

\usepackage{algorithm}
\usepackage{algorithmicx}
\usepackage{algpseudocode}
\usepackage[inline]{enumitem}

 \acmDOI{none}

\setcopyright{none}

\usepackage[textsize=tiny, textwidth=1.5cm]{todonotes}

\begin{document}
\setlength{\marginparwidth}{1.5cm}

\title{Privacy-Preserving Platform for Transactive Energy Systems}

\author{Karla Kvaternik}
\affiliation{\institution{Siemens Corporate Technology}}
\email{karla.kvaternik@siemens.com}

\author{Aron Laszka}
\affiliation{\institution{Vanderbilt University}}
\email{aron.laszka@vanderbilt.edu}

\author{Michael Walker}
\affiliation{\institution{Vanderbilt University}}
\email{michael.a.walker.1@vanderbilt.edu}

\author{Douglas Schmidt}
\affiliation{\institution{Vanderbilt University}}
\email{d.schmidt@vanderbilt.edu}

\author{Monika Sturm}
\affiliation{\institution{Siemens Corporate Technology}}
\email{monika.sturm@siemens.com}

\author{Martin Lehofer}
\affiliation{\institution{Siemens Corporate Technology}}
\email{martin.lehofer@siemens.com}

\author{Abhishek Dubey}
\affiliation{\institution{Vanderbilt University}}
\email{abhishek.dubey@vanderbilt.edu}


\vspace{-0.1in}

\renewcommand{\shortauthors}{K. Kvaternik et al.}

\begin{abstract}
Transactive energy systems (TES) are emerging as a transformative solution for the problems faced by distribution system operators due to an increase in the use of distributed energy resources and a rapid acceleration in renewable energy generation. These, on one hand, pose a decentralized power system controls problem, requiring strategic microgrid control to maintain stability for the community and for the utility. On the other hand, they require robust financial markets operating on distributed software platforms that preserve privacy. In this paper, we describe the implementation of a novel, blockchain-based transactive energy system. We outline the key requirements and motivation of this platform, describe the lessons learned, and provide a description of key architectural components of this system.
\end{abstract}


\maketitle

\vspace{-0.1in}
\section{Introduction}

\textbf{Emerging Trends:} Transactive energy systems (TES) have emerged as an anticipated outcome of the shift in electricity industry, away from centralized, monolithic business models characterized by bulk generation and one-way delivery, toward a decentralized model in which end users play a more active role in both production and consumption \cite{NIST_TE} \cite{Gridwise}. 
In this paper, we consider a class of TES that operates in grid-connected mode. The main actors are the consumers, which are comprised primarily of residential loads and prosumers who operate distributed energy resources (DERs), such as rooftop, solar batteries or flexible loads capable of demand/response. Additionally, a distribution system operator (DSO) manages the grid connection of the network. Such installations are equipped with an advanced metering infrastructure consisting of TE-enabled smart meters. In addition to the standard functionalities of smart meters: i.e. the ability to measure line voltages, power consumption and production, and communicate these to the distribution system operator (DSO); TE-enabled smart meters are capable of communicating with other smart meters, have substantial on-board computational resources, and are capable of accessing the Internet and cloud computing services as needed. Examples of such installations include the well-known Brooklyn Microgrid Project, \cite{BrooklynMicrogrid} and the Sterling Ranch learning community (currently under development) \cite{SterlingRanch}. A key component of TES is a transaction management platform (TMP), which handles all market clearing functions in a way that balances supply and demand in the local market.

\textbf{Why Blockchains?:} The capabilities of TE-enabled meters allow them to form a blockchain (BC) based TMP executing a market mechanism, using smart contracts \cite{Szabo97}. Examples of BC systems capable of executing smart contracts include Ethereum \cite{Buterin2013} and Hyperledger Fabric \cite{HyperledgerFabric16}.
There are a number of appealing properties of BC systems that motivate their use in a TMP. Firstly, BC technology enables the digital representation of energy and financial assets, and their secure transfer from one set of parties to another. By design, the security of this value transfer is guaranteed by the interaction protocol itself and obviates the need for trusted transaction intermediaries. Secondly, the execution of smart contracts (i.e. code that captures the market logic and participant roles) is automated and guaranteed.
Thirdly, the blockchain constitutes an immutable, complete, and fully auditable record of all transactions that have ever occurred in the BC system. These properties ensure market transparency, as well as the availability of a detailed market load profile, and grid utilization data. Thus, \cite{PowerLedger,TenneT,Lo3Patent} have already considered such implementations.


\textbf{Open challenges:} Existing initiatives such as \cite{PowerLedger,Lo3Patent,TenneT}
do not consider the impact of the electricity market on the controller responsible for the stability of the system due to the expectation that the bulk grid will maintain system stability. Furthermore, although these solutions present interesting case studies, they provide only a subset of services, which do not affect the overall power flow on the grid in a significant way. For example, they do not address the security, stability, and privacy requirements, which we describe in the next section. The information technology backbone that allows energy trades in an open P2P market to take place anonymously, and securely, has yet to be developed \cite{7747895,7725894}.

\textbf{Contributions $\rightarrow$ Design and Implementation of  \emph{Privacy-preserving
    Energy Transactions} (PETra)}
We focus on the specification of the components of the platform, their interfaces, and the distributed ledger in the system. Additionally, we highlight the architectural and protocol specifications of our platform, which ensure \emph{privacy} and \emph{security} for participants in the TES, as well as \emph{safety} for the TES. 
The specific contributions of this paper are (a) proposal of a set of TES requirements extracted from our experience, (b) architectural and protocol specification and implementation of a new blockchain-based middleware called PETra, first introduced in  \cite{Laszka17}, and (c) an empirical demonstration of the PETra functionality using an actual load profile data set from a microgrid test installation in Germany.  In particular, the high volumes of microtransactions in the envisioned TES pose challenges related to real-time communication of sensor data; for example, request-reply messaging between TMP modules, and other signaling that occurs outside of the blockchain. We will refer to these as ``out of band communications.''

The outline of this paper is as follows. Section \ref{sec:requirements} describes the requirements for this class of distributed systems. Section \ref{sec:related} describes the state of the art. We present our solution in section \ref{sec:petra}. It is followed by an evaluation and discussion in Sections \ref{sec:results} and \ref{sec:discussion}.



\section{Requirement Analysis}\label{sec:requirements}


The trading scenarios we consider involve consumers and prosumers that participate in a local P2P energy trading market by posting offers to sell produced energy, or offers to buy and consume energy in each consecutive time interval. An offer consists of quantity of energy being bought or sold, the time interval in which the trade is to be made, and possibly a reservation price - the maximum (or respectively, minimum) price at which the buyer (or respectively, seller) is willing to trade. The DSO bargains on the bulk market and provides all residual supply and demand within the microgrid. 

We assume that each participant has a means of predicting her future power production and consumption based on historical data, and does so prior to trading on the market. An example of a home energy management system that provides this means is the Siemens Energy IP Analytics Suite.
Moreover, each participant is represented by an automated trading agent that strategically posts offers to the TMP based on these predictions and the participant's personal trading goals.

In the simplest trading scenario, the DSO sets the price $p$ per kWH for the local market; $p$ is the price paid by any buyer and received by any seller, including the DSO.  The DSO can then dynamically adjust the price $p$ to affect the market efficiency, evaluated as the number of local transactions  vs. energy demand being met from a bulk supplier. Another scenario includes a fully dynamic market where all sellers, including the DSO, post offers that include a reservation price. Each consumer then picks a selling offer on a first-come, first-served basis. An extension of this scenario involves double auctions where both selling and buying offers are posted to the TMP, which executes an automated, regulator-approved market clearing algorithm as an immutable smart contract on the TMP's blockchain system. This algorithm selects the clearing price $p$ within each time interval. With respect to these trading scenarios we propose the following requirements.

\textbf{Communication Fabric}
The first requirement is the existence of an appropriate communication and messaging architecture. The TMP must collect participants' offers and make them available to buyers, and the market algorithm must communicate clearing prices, buyer-seller matchings, or other market-related signals depending on the trading scenario. In order to meet the operational and safety requirements described next, these messages must be reliably delivered under strict timing constraints, derived from the deadline by which a trade must clear. Moreover, the TMP must be capable of handling high volumes of micro-transactions anticipated in P2P trading scenarios.  Finally, the communication fabric must support confidentiality, integrity, and non-repudiation of transactional data.


\textbf{Operational Safety and Cyber-Physical Security}
The trading activity permitted by the TMP shall not compromise the stability of the physical system operation. Moreover, congestion constraints along any feeder shall be respected.\footnote{In the context of grid-connected microgrids, system stability refers to real-time balancing -- i.e. the system's ability to dynamically match supply and demand as closely as possible, and a tendency to drive the difference between supply and demand to zero under small perturbations. Resiliency refers to the system's ability to react to contingencies and recover from faults. Congestion on a transmission line occurs when the power flow exceeds the line's maximum rated capacity.} This also requires assurance that malicious or negligent trading activity is discouraged.
\footnote{Negligent trading may include producers who commit to a certain production level and fail to deliver. Transactional security means that the execution of contractual obligations among all participants, including the DSO, is guaranteed.} 
Finally, the TMP should have provisions for preventing or detecting negligent or malicious interference with smart meters - i.e. the adversarial or natural attacks against the interface between the physical world and the blockchain; data logged shall be securely communicated to the DSO and requests made by the meter on behalf of the prosumer shall be accurately recorded on the blockchain. 

\textbf{Market Security}
The TMP shall include provisions for ensuring the protection of consumer interests, as well as those of the DSO. Consumer interests include being billed correctly and fairly
based on energy prices set by the DSO and the measurements made by the smart meters. Additionally, it is important to ensure all prosumers will be allowed to participate in the market fairly.

\textbf{Privacy}
Information such as the amount of energy produced, consumed, bought, or sold by any prosumer should be available only to the DSO and the essential market functions of the TMP. All bids and asks, and the matching thereof, should remain anonymous. A participant's energy usage patterns and personal information, such as financial standing, shall not be inferable from the participant's trading activity\footnote{Inference of energy usage patterns can be exploited by inferring the presence or absence of a person in their home, for example.}.






\section{Analysis of State of the Art}
\label{sec:related}
The TMP system requires peer-to-peer messaging, enabling each stakeholder to receive all the required `bid' messages, concerning a specific ask. 
Thereafter, a consumer can choose to accept a bid and inform the ledger about the acceptance.  
Once the bid is accepted, the transaction is recorded into a distributed ledger in a way that allows everyone in the community to agree that the transaction took place. Once consensus is established, the transaction is deemed successful, and we  say that the market has cleared. In the context of this workflow, we next describe the state of the art across the two dimensions of Application and Communication platforms in smart grid and distributed transaction management platform for smart grid.

\textbf{Application Platforms for Smart Grid} There seem to be two approaches in general for moving power applications
from centralized to distributed processing paradigms.
One approach is to consider each remote computing entity (or node) as an \textit{agent} \cite{underfreq-2015} or \textit{actor} \cite{agha1985actors} \cite{LeeNiddodiSrivastavaBakken2016} that communicates via messages with other agents or actors, and focuses on specific grid issues such as state estimation, remedial action schemes, and load shedding. The other approach utilizes each remote computing entity as an open application platform that can host multiple applications managing varied aspects of the local grid  \cite{freedmSysArch2014}. Both approaches utilize messaging between nodes, and leverage a common set of services on each node, to handle distributed coordination concerns. \cite{LeeNiddodiSrivastavaBakken2016} calls for group membership, leader election, voting, group consensus or agreement on data values, mutual exclusion on access to shared resources, and multicast communication with same order and atomic properties. Both \cite{underfreq-2015} and \cite{LeeNiddodiSrivastavaBakken2016} prototype their approaches using MATLAB toolkits, with \cite{LeeNiddodiSrivastavaBakken2016} utilizing the Akka Java toolkit to model actors. \cite{Meng2010} developed simulations using the SimPowerSystem software in the Simulink environment. Our application platform, called Resilient Information Architecture Platform for Smart Grid (RIAPS)
\cite{riaps1}, provides actor and component based abstraction, as well as support for deploying algorithms on devices across the network\footnote{RIAPS uses ZeroMQ \cite{hintjens2010zeromq}, and Cap'n Proto \cite{varda2015cap}, to manage the communication layer.} and solves problems collaboratively by providing micro-second level time synchronization \cite{riaps2}, failure based reconfiguration \cite{dubey2017resilience}, and group creation and coordination services (still under active development), in addition to the services described in \cite{LeeNiddodiSrivastavaBakken2016}. It is capable of handling different communication and running implemented algorithms in real-time.

\textbf{Transaction Management Platforms (TMP) for Smart grid} 
TMP require communication, as well as trading mechanisms that provide the capability to match the bids and asks. Additionally, they must provide fairness and integrity assurances.  Blockchain based solutions have the potential to enable large-scale energy trading based on distributed consensus systems. However, popular blockchain solutions, such as Bitcoin \cite{Satoshi} and Ethereum \cite{buterin2013ethereum} suffer from design limitations that prevent their direct application to validating energy trades. In particular, their transaction-confirmation time is relatively slow and variable, primarily due to the proof-of-work algorithm and most of the communication occurring via the ledger.
For example, Aitzhan and Svetinovic implemented a proof-of-concept platform for decentralized smart grid energy trading using blockchains, but their system is based on proof-of-work consensus, and they do not consider grid control and stability, or scalability~\cite{aitzhan2016security}. Additionally, there is the problem of privacy - all transactions in these systems are  public \cite{kosba2016hawk}. 

Most works in this area have focused on the privacy issue from the context of smart meters. McDaniel and McLaughlin discuss the
privacy concerns of energy usage profiling, which smart grids could
potentially enable~\cite{mcdaniel2009security}. Efthymiou and Kalogridis describe a method for securely anonymizing frequent electrical metering data sent by a smart
meter~\cite{efthymiou2010smart} by using a third party escrow mechanism. Tan et
al.\ study privacy in a smart metering system from an information
theoretic perspective in the presence of energy harvesting and storage
units~\cite{tan2013increasing}. They show that energy harvesting
provides increased privacy by diversifying the energy source, while a
storage device can be used to increase both energy efficiency and
privacy. However, the transaction data provides more fine-grained information than the smart meter usage patterns \cite{Privacy2017}. 

PETra extends these works by (1) leveraging a decentralized computation fabric provided by smart homes in the community,
(2) addressing the privacy
threat posed by trading using a novel trading sequence implementation, (3) showing how partial trades can be fulfilled, and (4) using off-blockchain communication primitives provided by the distributed application management platform RIAPS. While the conceptual design of PETra was presented in \cite{Laszka17}, this paper describes the revised protocol and the trading algorithm, and presents the implementation results. 

\section{Our Solution - PETra}

\label{sec:petra}

\begin{figure}
\resizebox {0.7\columnwidth} {!} {
\begin{tikzpicture}[x=1.5cm, y=1.5cm, font=\small,
  Component/.style={fill=white, draw, align=center, rounded corners=0.1cm, drop shadow={shadow xshift=0.05cm, shadow yshift=-0.05cm, fill=black}},
  Connection/.style={<->, >=stealth, shorten <=0.05cm, shorten >=0.05cm}]

\foreach \pos/\name in {0/pros1, 0.8/pros2, 1.6/pros3} {
  \node [Component] (\name) at (\pos - 4, \pos) {Prosumer\\(\texttt{RIAPS, geth})};
}

\node [Component] (dso) at (-1, 2.6) {DSO\\(\texttt{RIAPS, geth})};

\fill [fill=black!15] (90:1.5) -- (200:1.5) -- (340:1.5) -- (90:1.5);

\foreach \pos in {90, 200, 340} {
  \node [Component] at (\pos:1.5) {Ethereum\\miner (\texttt{geth})};
}

\node [Component, dotted] (contract) at (0, 0) {Smart contract\\(\texttt{Solidity})};

\draw [Connection, bend left=65] (pros1) to node [midway, left, shift={(-0.25,0)}] {RIAPS ($\emptyset$MQ)} (dso);
\draw [Connection, bend left=55] (pros2) to (dso);
\draw [Connection, bend left=45] (pros3) to (dso);

\draw [Connection, bend right=15] (dso) to (contract);
\draw [Connection, bend right=0] (pros1) to node [midway, below left] {Ethereum} (contract);
\draw [Connection, bend right=0] (pros2) to (contract);
\draw [Connection, bend right=0] (pros3) to (contract);
\end{tikzpicture}

}
\vspace{-0.1in}
\caption{Components of PETra. The DSO and prosumers are comprised of RIAPS components and \texttt{geth} Ethereum clients. The smart contract is implemented in Solidity, a high-level language for Ethereum, and it is executed by a network of \texttt{geth} miners.}
\vspace{-0.1in}
\label{fig:components}
\end{figure}
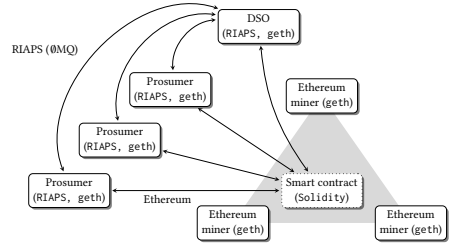

\begin{figure*} [ht]

\center

\begin{tikzpicture}[x=5.3cm, y=-0.068cm, font=\small,
RedDashed/.style={red, dashed}]
\def\posDSO{0}
\def\posProsumer{1}
\def\posLedger{2}
\def\posOtherProsumer{3}
\foreach \name/\pos in {DSO/\posDSO, Prosumer/\posProsumer, Smart contract/\posLedger, Other prosumer/\posOtherProsumer} {
  \draw (\pos, 0) -- (\pos, 91);
  \node [draw, fill=white, rounded corners=0.1cm] at (\pos, 0) {\name};
}
\newcounter{seqTime}
\setcounter{seqTime}{0}
\foreach \action/\from/\to/\styl/\delta in {
  {withdrawAssets(anonAddress, assets)/\posProsumer/\posDSO/solid/11}, 
  {failedWithdrawal(anonAddress, msg)/\posDSO/\posProsumer/red/8},
  {addEnergyAsset(anonAddress, asset), addFinancialBalance(anonAddress, amount)/\posDSO/\posLedger/solid/8},
  {AssetAdded(anonAddress, assetID, asset)/\posLedger/\posProsumer/dashed/8},
  {postOffer({assetID, price})/\posProsumer/\posLedger/solid/8},
  {OfferPosted({offerID, assetID, price})/\posLedger/\posOtherProsumer/dashed/4},
  {rescindOffer(offerID)/\posProsumer/\posLedger/red/6},
  {OfferRescinded(offerID)/\posLedger/\posOtherProsumer/RedDashed/4},
  {acceptOffer({offerID, assetID})/\posOtherProsumer/\posLedger/solid/8},
  {OfferAccepted({offerID, assetID})/\posLedger/\posProsumer/dashed/4},
  {depositEnergyAsset(assetID), depositFinancial(amount)/\posProsumer/\posLedger/solid/14},
  {AssetDeposited(anonAddress, asset), FinancialDeposited(anonAddress, amount)/\posLedger/\posDSO/dashed/8}%
} {
  \addtocounter{seqTime}{\delta}
  \draw [->, >=stealth, shorten <=0.05cm, shorten >=0.05cm, \styl] (\from, \value{seqTime}) -- (\to, \value{seqTime}) node [midway, above, align=center, text width={abs(\to - \from) * 4.8cm}, fill=white, fill opacity=0.67, text opacity=1] {\footnotesize\texttt{\action}};
}

\end{tikzpicture}
\vspace{-0.2em}
\caption{Sequence diagram of the trading workflow. Solid lines represent RIAPS messages and Ethereum transactions, while dashed lines represent smart-contract events. Messages and transaction in \textcolor{red}{red} stop the trading workflow.}
\label{fig:workflow}
\vspace{-0.12in}
\end{figure*}
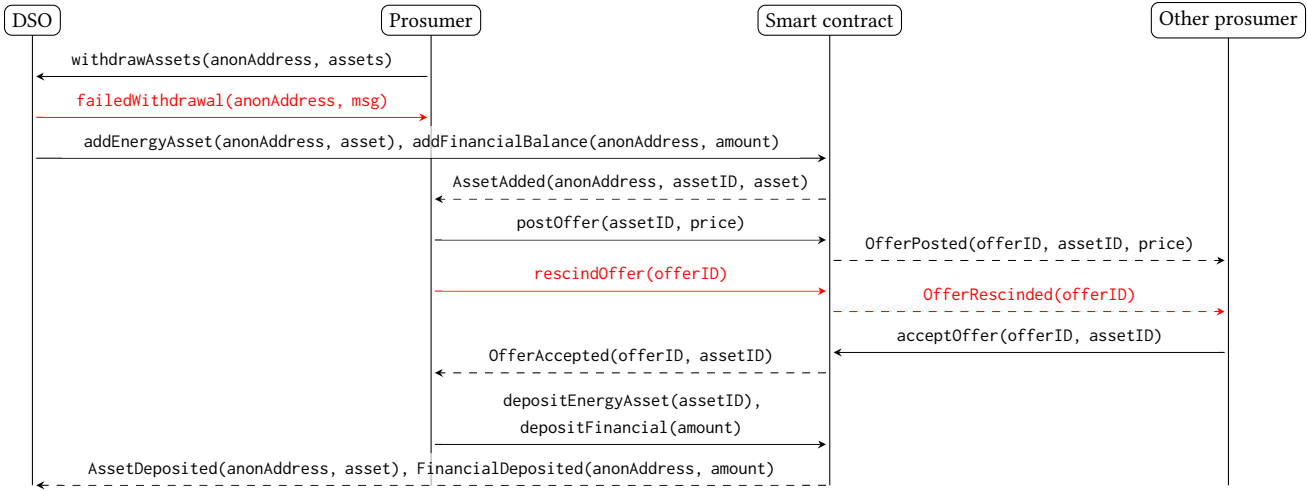

Our system contains the following types of components (see Figure~\ref{fig:components} for an illustration):
\begin{itemize}[leftmargin=*]
\setlength{\itemsep}{0pt}%
    \setlength{\topsep}{0pt} 
    \setlength{\partopsep}{0pt}
    \setlength{\parsep}{0pt}
    \setlength{\parskip}{0pt}%
\item DSO: There is a single component of this type, which represents the Distribution System Operator of the microgrid. 
The primary responsibilities of this component are ensuring the safe operation of the microgrid and regulating the total load of the microgrid.
To this end, the DSO component can limit the energy and financial assets that the prosumers' withdraw for trading, and it can also set a price policy for the microgrid.
Note that this component does not have to be online during trading, so the reliability of the system does not hinge on the reliability of this component.
\item Prosumer: There is a component of this type for every household.
The prosumer components are responsible for trading energy production and consumption for their households.
To do so, a component first estimates the future production and consumption of the household, withdraws energy production or consumption assets from the DSO, and then trades these assets with other prosumers.
To ensure that trading does not compromise the household's privacy, the component uses randomly generated anonymous addresses for trading, which hide the identities of trade partners from each other.
\item Smart contract: 
This component (deployed as an Ethereum contract on the private blockchain) is responsible for keeping track of the energy and financial assets belonging to each anonymous address, enabling prosumers to post trade offers, and exchanging assets when another prosumer decides to take an offer.
The contract is executed in a decentralized manner by a network of miners, which provides reliability. Additionally, we have several Ethereum clients, one per prosumer and one for the DSO, which interact with the smart contract.
\end{itemize}

\vspace{-0.08in}
\subsection{Assets and Data Structures}

The ability to specify points or intervals in time is crucial.  For
example, control signals specify how the microgrid load should change
at certain points in time, energy trades specify when energy will be
consumed or produced, etc.  To facilitate representing signals and
transactions, we divide time into fixed-length intervals, and specify
points or periods in time using these discrete timesteps.  The length
of the time interval is determined based on the timing assumptions of
the physical power system.  For example, the
time interval may as low as 4 seconds, which corresponds to how frequently
the control signal of the DSO typically changes \cite{federal2011frequency}.

Prosumers trade energy production and consumption with each other, which are represented in PETra by energy assets, which is a structure that comprises the following fields: (a)
 \texttt{int64 power}: non-negative amount of power to be produced or consumed (for example, measured in watts), (b)
 \texttt{uint64 start}: first time interval in which energy is to be produced (or consumed), and (c) \texttt{uint64 end}: last time interval in which energy is to be produced (or consumed).
An asset with positive power value represents energy production, and we call it an \texttt{EnergyProductionAsset}.
An asset with negative power value, on the other hand, represents energy consumption, and we call it an \texttt{EnergyConsumptionAsset}.

Energy trading must also involve the transfer of currencies, which are represented by financial assets.
A \texttt{FinancialAsset} is simply an \texttt{uint64} value, denominated in a fiat currency.

\subsection{Trading Workflow}

Next, we discuss the trading workflow that is used by prosumers to trade energy production and consumption assets, as well as financial assets with each other.
This workflow involves both off-blockchain messaging (using RIAPS), and on-blockchain transactions and events.
We list these messages, transactions, and events in the order in which they typically appear in the workflow.
Figure~\ref{fig:workflow} shows a graphical illustration of the workflow.
\vspace{-0.1em}
\begin{itemize}[leftmargin=*]
\setlength{\itemsep}{0pt}%
    \setlength{\topsep}{0pt} 
    \setlength{\partopsep}{0pt}
    \setlength{\parsep}{0pt}
    \setlength{\parskip}{0pt}%
\item 
\texttt{withdrawAssets(anonAddress, assets)}: RIAPS message sent by a prosumer to the DSO, asking the DSO to transfer energy and/or financial assets from the prosumer's account at the DSO to an anonymous address to protect her privacy.
Before sending this message, the prosumer should generate a new random anonymous address.
The message specifies the assets that the prosumer wishes to withdraw, and the anonymous address to which the DSO should transfer them, which must be cryptographically signed by the prosumer.
Note that the prosumer may send this message long before actually engaging in trading, so the DSO does not have to be online continuously.
\item \texttt{failedWithdrawal(anonAddress, msg)}: RIAPS message sent by the DSO to the prosumer,
notifying the prosumer that the requested assets cannot be withdrawn due to, e.g., energy safety requirements or insufficient funds.
\item \texttt{addEnergyAsset(anonAddress, asset)},\\\texttt{addFinancialBalance(anonAddress, amount)}:
smart contract transaction called by the DSO, creating energy and financial assets on the blockchain and transferring them to an anonymous address.
Before recording this transaction, the DSO must first verify whether enabling the prosumer to trade these assets would violate any safety requirements.
The transaction specifies the assets and the anonymous address to which they are transferred, and it must be cryptographically signed by the DSO.
\item \texttt{AssetAdded(anonAddress, assetID, asset)},\\\texttt{FinancialAdded(anonAddress, amount)}: events broadcast by the smart contract,
notifying the prosumer that the requested assets have been transferred to the anonymous address.
\item \texttt{postOffer(assetID, price)}: smart contract transaction called by a prosumer, publicly posting an energy bid or ask.
If the prosumer is interested in buying energy, then it posts an energy bid, which specifies an energy consumption asset and a price.
If the prosumer is interested in selling, then it posts an energy ask, which specifies an energy production asset and a price.
vIn both cases, the transaction must be cryptographically signed by the private key of the address, and it locks the assets until the offer is accepted or rescinded.
\item \texttt{OfferPosted(offerID, assetID, price)}:
event broadcast by the smart contract, notifying prosumers that an offer was posted.
\item \texttt{rescindOffer(offerID)}:
smart contract transaction called by a prosumer, rescinding an offer.
The transaction must be cryptographically signed by the private key of the poster.
\item \texttt{acceptOffer(offerID, assetID)}:
smart contract transaction called by a prosumer,
accepting a previously posted offer. 
If the offer was an energy bid, then the other prosumer has to provide an energy production assets;
if the offer was an energy ask, then the other prosumer has to provide both energy consumption and financial assets.
In both cases, the transaction must be cryptographically signed by the private key of the other prosumer's anonymous address.
If there is an overlap between the time intervals of the offered asset, and the asset provided by the other prosumer, then the intersecting parts of the assets are exchanged and the non-overlapping parts are returned to their original owners.
Similarly, based on the price and exchanged energy assets, a part of the financial asset is transferred to the seller, while the rest is returned to the seller. 
\item \texttt{OfferAccepted(offerID, assetID)}:
event broadcast by the smart contract,
notifying the prosumer that its offer has been accepted, and the assets have been exchanged.
\item \texttt{depositEnergyAsset(assetID)},\\\texttt{depositFinancial(amount)}:
smart contract transactions called by a prosumer,
depositing energy and financial assets to the prosumer's account.
The transaction specifies the assets, and it must be cryptographically signed by the anonymous address that owns them.
Note that to protect privacy, the transaction does not specify the prosumer, so the DSO has to keep track of which prosumer has used which anonymous address.
\item \texttt{AssetDeposited(anonAddress, assetID)},\\\texttt{FinancialDeposited(anonAddress, amount)}:
event broadcast by the smart contract,
notifying the DSO that assets have been deposited from anonymous address, which triggers the transfer of these assets to the prosumer's account at the DSO.
\end{itemize}

\subsection{Case Study}
\label{sec:results}

\definecolor{blueLine}{RGB}{57,106,177}
\definecolor{blueFill}{RGB}{114,147,203}
\definecolor{redLine}{RGB}{204,37,41}

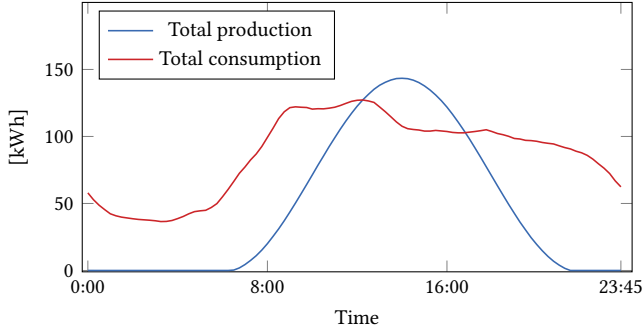
\begin{figure}[t]
\centering

\begin{tikzpicture}
\begin{axis}[
  font=\small,
  width=1.05\columnwidth,
  height=0.61\columnwidth,
  ymin=-1,
  ymax=200,
  xmin=-1,
  xmax=97,
  legend pos=north west,
  xlabel=Time,
  ylabel={[kWh]},
  ytick={0, 50, 100, 150},
  xtick={0, 32, 64, 95},
  xticklabels={0:00, 8:00, 16:00,  23:45},
]
\addplot[no markers, solid, blueLine, semithick] table[x expr=\coordindex, y=Production, comment chars={\%}] {total_prod_cons.csv};
\addlegendentry{Total production};
\addplot[no markers, solid, redLine, semithick] table[x expr=\coordindex, y=Consumption, comment chars={\%}] {total_prod_cons.csv};
\addlegendentry{Total consumption};
\end{axis}
\end{tikzpicture}
\vspace{-2em}
\caption{Load profile and Generation Profile in KWH per 15 minute interval. The horizontal axis shows time of day.}
\label{fig:profile}
\end{figure}

We use  data collected by Siemens, from a microgrid in Germany,  to demonstrate a simulated transactive scenario. Figure \ref{fig:profile} shows the total energy produced in this system over the day, and  the total energy consumed. We use a $T=15$ minute time interval for bids and asks. We picked a 3.5 hour time interval (from 2:15pm to 5:45pm) and 2 producers and 7 consumers that overlapped with the peak of production capacity (see Figure \ref{fig:profile}). We ran the network across six virtual machines, each with 2 virtual CPUs, 4 GB RAM and 40 GB hard-disk. The \texttt{geth} clients (one per actor) and miners were equally distributed on this network. The actors (Prosumers and DSO) were written in Python, and they communicated with the \texttt{geth} clients using JSON-RPC API provided by Ethereum. The  actors communicated with each other using RIAPS and polled the blockchain ledger for transactional updates using custom filters, which are supported by the Ethereum API.

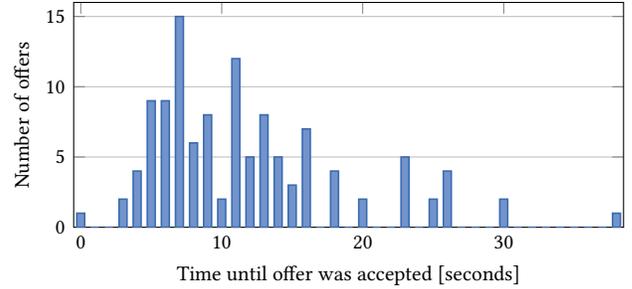
\begin{figure}[t]
  \begin{tikzpicture}
\begin{axis}[
  font=\small,
  width=1.05\columnwidth,
  height=0.54\columnwidth,
  ymin=0,
  ymax=16,
  xmin=-0.5,
  xmax=38.5,
  xlabel={Time until offer was accepted [seconds]},
  ylabel={Number of offers},
  ymajorgrids,
]
\addplot[ybar, bar width=3pt, no markers, fill=blueFill, draw=blueLine, semithick] table[x=Seconds, y=Count, comment chars={\%}, col sep=comma] {TimePerOfferHistogram.csv};
\end{axis}
\end{tikzpicture}
\vspace{-1em}
\caption{Histogram of time it takes to clear the two transactions related to post offer and accept offer. 90\% of the trades were closed within 23 seconds or less.}
\label{fig:offer_histogram}
\end{figure}

Figure~\ref{fig:offer_histogram} shows the distribution of the time between when an offer was made by a producer, and then the time when the offer was accepted by a consumer and cleared. As shown by Figure~\ref{fig:workflow}, this includes two transactions, \texttt{postOffer} and \texttt{acceptOffer}, which have to be verified and recorded by the miners. To clear these two transactions, at least two blocks need to be mined. The statistics of the clearing-time distribution are as follows: 
average = 11.79 seconds, median = 11 seconds, variance = 46.74, maximum = 38 seconds, minimum = 0 seconds, and 90\% of trades were cleared within 23 seconds or less.


\subsection{Requirements Analysis Discussion}
 \label{sec:discussion} 
We conclude this section with a brief discussion of how PETra satisfies the requirements outlined in Section~\ref{sec:requirements}.

\textbf{Communication Fabric:}
The key requirements for communication are reliability and security. 
In PETra, communication and messaging services are built on (a) RIAPS between the DSO and prosumers, and (b) blockchain transactions and events between the smart contract and other components.
The RIAPS communication layer \cite{eisele2017riaps} presents a reliable messaging service, which  is being currently extended to provide message confidentiality, integrity, and non-repudiation with the help of digital signatures.
Since communication between prosumers and the DSO (i.e. withdrawal) may happen well in advance of actual trading, the DSO and the messaging service do not have to be online continuously.
Combined with the features of RIAPS, this flexibility in uptime leads to a very high level of reliability. 

For messaging between the smart contract and other components, the blockchain provides a secure and reliable communication medium.
The  blockchain ledger is an immutable, complete, and fully auditable record, which guarantees integrity and non-repudiation for transactions and events.
Note that---by design---the blockchain does not provide confidentiality, since every transaction and event is public; we will discuss privacy implications and requirements in detail below.
Finally, the blockchain provides a high level of reliability since the ledger is maintained by multiple nodes, which can reach consensus even in the presence of some misbehaving or malicious nodes.

\textbf{Operational Safety, Cyber-Physical Security, and Market Safety:}
For safety and cyber-physical security, it is crucial to ensure that  trading activity cannot compromise the stability of the grid and congestion constraints are respected.
PETra achieves these goals by enabling the DSO to tightly control the amount of energy that a prosumer may (offer to) sell or buy.
A prosumer's energy trading workflow (see Figure~\ref{fig:workflow}) always begins with a withdrawal from the DSO.
By limiting the amount assets that can be withdrawn, the DSO limits the bids and asks that may be posted by a prosumer, thereby enforcing safety requirements (e.g., preventing a prosumer from offering to produce more power than her production capacity).
Fine-grained withdrawal rules based on time, power, etc, can be used to prevent a wide range of negligent or malicious trading.

To protect the prosumers' interests, 
we must enable them to detect and prove if they are incorrectly billed or denied fair participation in the market.
PETra meets these goals due to the public, fully auditable, and immutable nature of the blockchain ledger.

\textbf{Privacy:}
Privacy requirements dictate that prosumers cannot gain information regarding other prosumers' consumption and production---not even if they are trade partners.
This requirement presents an interesting challenge since every transaction on the blockchain ledger is public.
PETra provides privacy through pseudonymous trading; instead of real identities, prosumers use randomly chosen addresses for trading with each other.
However, pseudonymous addresses could be de-anonymized either by (a) learning which addresses belong to the same prosumer or (b) using the prosumers' communication addresses (e.g., IP addresses used to send transactions).
Firstly, by employing a large number of anonymous addresses, a prosumer can effectively prevent de-anonymization attacks that would link her addresses together.\footnote{Note that generating new addresses is trivial.} 
Secondly, by combining our platform with a communication anonymity solution, such as onion routing, we can prevent de-anonymization based on communication addresses.






\section{Conclusions and Future Work}
A transaction management platform (TMP) is the key component of a transactive energy system. The role of the TMP is to facilitate the deployment of applications that help maintain stability of the microgrid, as well as implement efficient market mechanisms and enable an open P2P energy trading market. In this paper we proposed a blockchain-based TMP called PETra, which extends existing works by (1) leveraging a decentralized computation
fabric provided by the smart homes in the microgrid, (2) addressing the privacy threat posed by trading using a novel trading sequence implementation, (3) showing how partial trades can be fulfilled, and (4) using off-blockchain communication primitives provided by the distributed application management platform RIAPS. 

During the experiments, we made the following observations:
Blockchains are not enough by themselves to implement a full-fledged TMP for transactive energy systems. We need off-blockchain communication for (1) performance (transactions may be slow) (2) reliability (transactions may be lost before they are permanently recorded) (3) privacy (we can anonymize assets by mixing on the blockchain, but doing it off-blockchain is much more efficient). Additionally,  existing smart contract languages (e.g., Solidity) have some serious limitations (and peculiarities), which complicate the implementation of some domain logic. For instance, at the time of writing, Solidity did not provide floating-point data types.

\bibliographystyle{ACM-Reference-Format}
\bibliography{references} 

\end{document}